# DYNAMIC QUANTIZED FRACTURE MECHANICS

N. M. Pugno[*] and R. S. Ruoff[#]

[*]Department of Structural Engineering, Politecnico di Torino, Corso Duca degli Abruzzi 24, 10129, Torino, Italy; nicola.pugno@polito.it; n-pugno@northwestern.edu

[#]Department of Mechanical Engineering, Northwestern University, Evanston, IL 60208-3111, USA; r-ruoff@northwestern.edu

**Abstract**

A new quantum action-based theory, Dynamic Quantized Fracture Mechanics (DQFM), is presented that modifies continuum-based dynamic fracture mechanics. The crack propagation is assumed as quantized in both space and time. The static limit case corresponds to Quantized Fracture Mechanics (QFM), that we have recently developed to predict the strength of nanostructures. Here, we discuss the case of fracture strength of carbon nanotubes and show that, in contrast to the conclusion reported in a recently published article (that "materials become insensitive to flaws at nanoscale") even a single atomic vacancy causes a significant reduction in strength.

DQFM predicts the well-known forbidden strength and crack speed bands–observed in atomistic simulations–which are unexplained by continuum-based approaches. In contrast to Dynamic Fracture Mechanics (DFM) and Linear Elastic Fracture Mechanics (LEFM) that are shown to be limiting cases of DQFM and which can treat only *large* (with respect to the "fracture quantum") and sharp cracks under moderate loading speed, DQFM has no restrictions on treating defect size and shape, or loading rates. Simple examples are discussed: (i) strengths predicted by DQFM for static loads are compared with experimental and numerical results on carbon nanotubes containing nanoscale defects; (ii) the dynamic fracture initiation toughness predicted by DQFM is compared with experimental results on microsecond range dynamic failures of 2024-T3 aircraft aluminum alloy.

An analogy between DQFM and quantum mechanics, both based on action quanta, is presented. The strength of the carbon nanotube-based cable for the *Space Elevator* is also discussed. Since LEFM has been successfully applied also at the geophysics size-scale, it is conceivable that DQFM theory can treat objects that span at least 15 orders of magnitude in size.

1. Introduction

Two classic treatments of Linear Elastic Fracture Mechanics (LEFM) are Griffith's criterion [1], an energy-based method, and a method based on the stress-intensity factor developed by Westergaard [2]. These have been shown to be equivalent, as in the correlation between (static) energy release rate and stress-intensity factors formulated by Irwin [3]. An extension towards Dynamic Fracture Mechanics (DFM) was proposed by Mott [4], which included in Griffith's energy balance the contribution of the kinetic energy. Dynamic stress-intensity factors were then also proposed, as well as the dynamic generalization of Irwin's correlation, see the Freund's book [5]. Since LEFM and DFM can be applied only to large and sharp cracks under moderate loading rates, we choose to modify them by accounting for the discontinuous nature of matter and crack propagation, in both space and time.

Considering a balance of action quanta during crack propagation results in a more flexible theory without *ad hoc* assumptions. We call this *Dynamics Quantized Fracture Mechanics* (DQFM; note: we use the term "quantized" -as introduced by Novozhilov- not "quantum", that could be erroneously linked to Quantum Mechanics). Forbidden strength and crack speed bands clearly emerge. As *Quantized Fracture Mechanics* (QFM) [6], allows one to predict the strength of defective structures under quasi-static loading, DQFM can predict the strength (and the time to failure) under dynamic loading (and also the crack tip evolution). A comparison between QFM and theoretical/experimental/numerical investigations on fracture strength of carbon nanotubes, and between DQFM and experimental data on the dynamics fracture toughness of 2024-T3 aircraft aluminum alloy, is presented.

A considerable body of literature on fracture in discrete lattices has been developed over the past 25 years. In particular, the earliest work of its kind was probably by Slepyan [7]. This was followed by a number of important advances [8,9], summarized in a very complete work [10]. Additionally, researchers have published a number of related papers [11-13], see also the references quoted in [7-13]. Even if all these papers presented important ideas, we believe that our theory still represent a new contribute in this area, as a natural extension of dynamic fracture mechanics and quantized fracture mechanics.

## 2. Dynamic Fracture Mechanics

According to the principle of conservation of energy, during crack propagation the total energy (sum of the potential $W$, kinetic $T$, and dissipated $\Omega$, energies) is a constant. Thus, $\frac{\partial}{\partial A}(W+T+\Omega)=0$, where $\frac{\partial \Omega}{\partial A}=G_{dC}$ is the dynamic fracture energy (dissipated per unit area created) of the material. The dynamic energy release rate is defined as $G_d = -\partial(W+T)/\partial A$. The quasi-static condition refers to $T \approx 0$; thus, it is simply $G=G_C$, with $G=-\partial W/\partial A$ the (static) energy release rate and $G_C$ the (static) fracture energy. This represents the well-known Griffith's criterion, used for predicting the strength of cracked structures under quasi-static external loads. On the other hand, $G_d = G_{dC}$ allows one to consider dynamic external loads for predicting strength and time to failure and to describe the evolution of the crack tip. Let us assume a crack of length $l$ and speed $v=dl/dt$, where $t$ is time. From DFM, for a significant family of problems (see [5] for details), it is expected that $G_d(l,t,v)=g(v)G_d(l,t,0)$, where $g(v)$ is a universal function of the crack tip speed (see [5]):

$$g(v) \approx 1 - v/c_R, \qquad (1)$$

with $c_R$ Rayleigh speed. Introducing the function $g_C$ as:

$$g_C = \frac{G_{dC}}{G_C}. \qquad (2)$$

the simplest assumption corresponds to $g_C \approx 1$.

The Irwin correlation ($G = \frac{K_I^2}{E'} + \frac{K_{II}^2}{E'} + \frac{1+\nu}{E}K_{III}^2$) connects the (static) stress-intensity factor $K_{I,II,III}$ for opening (I), sliding (II), and tearing (III) crack propagation Modes with the (static) energy release rate $G$, through the elastic constants of the material ($E' = E$ for plane stress, or $E' = E/(1-\nu^2)$ for plane strain, where $E$ is Young's modulus and $\nu$ is the Poisson's ratio of the material). The extension in the dynamic regime yields the dynamic Irwin's correlation as $G_d = \frac{A_I(v)K_{dI}^2}{E'} + \frac{A_{II}(v)K_{dII}^2}{E'} + \frac{1+\nu}{E}A_{III}(v)K_{dIII}^2$, where $K_{dI,II,III}$ are the dynamic stress-intensity factors and $A_{I,II,III}(v)$ are universal functions of the crack tip speed $v$ (see [5]). In addition, from DFM, for a significant family of problems (see [5] for details), it is expected $K_{dI,II,II}(l,t,v) = k_{I,II,III}(v)K_{dI,II,III}(l,t,0)$, where $k_{I,II,III}(v)$ are again universal functions of the crack tip speed $v$ (see [5]):

$$k_I(v) \approx \frac{1-v/c_R}{\sqrt{1-v/c_D}}, \quad k_{II}(v) \approx \frac{1-v/c_R}{\sqrt{1-v/c_S}}, \quad k_{III}(v) \approx 1-v/c_S, \qquad (3)$$

where $c_D = \sqrt{\frac{E}{\rho}\frac{1-\nu}{(1+\nu)(1-2\nu)}}$ and $c_S = \sqrt{\frac{E}{\rho}\frac{1}{2(1+\nu)}}$ are the longitudinal and shear wave speeds respectively and $\rho$ is the material density ($c_R \approx 0.9c_S$). Thus, the dynamic Irwin's correlation implies in general:

$$A_{I,II,III}(v) = \frac{g(v)}{k_{I,II,III}^2(v)}. \qquad (4)$$

By rearranging the previous formulas one derives the condition for the incipient crack propagation in the quasi-static regime ($T \approx 0$) in the stress-intensity factor based treatment, i.e., $K_{I,II,III} = K_{I,II,IIIC}$, where $K_{I,II,IIIC}$ are the (static) critical stress-intensity factors or alternatively called the (static) fracture toughness. In dynamics the previous relation becomes $K_{dI,II,III} = K_{dI,II,IIIC}$, where $K_{dI,II,III}$ are the dynamic stress-intensity factors and $K_{dI,II,IIIC}$ represent the dynamic critical stress-intensity factors, or the dynamic fracture toughness. Note that to distinguish between $K_{dI,II,IIIC}(v)$ and $K_{dI,II,IIIC}(v=0)$ the former is called the dynamic fracture propagation toughness and the latter the dynamic fracture initiation toughness; in the same manner, $G_{dC}(v)$ and $G_{dC}(v=0)$ are the dynamic fracture propagation and initiation energies.

Defining the functions $k_{I,II,IIIC}$ as:

$$K_{dI,II,IIIC} = k_{I,II,IIIC} K_{I,II,IIIC}, \qquad (5)$$

for consistency with the energy balance it must be true that:

$$\frac{g_C}{k_{I,II,IIIC}^2} = \frac{g(v)}{k_{I,II,III}^2(v)} = A_{I,II,III}(v). \qquad (6)$$

We expect $g_C = g_C(v)$ and $k_{I,II,IIIC} = k_{I,II,IIIC}(v)$ and for $v$=0 $g_C = k_{I,II,IIIC} = 1$. According to eq. (6) the dynamic fracture propagation toughness and energy cannot be considered both coincident with their initiation values since $g_C = k_{I,II,IIIC} = 1$ cannot be satisfied for $v \neq 0$. In addition, at the incipient crack propagation the dynamic fracture initiation toughness and energy should be identical to their static values. The experiments in general do not agree with this result; however we will show this to be a consequence of adopting the classical criterion rather than due to the real nature of materials.

## 3. Dynamic Quantized Fracture Mechanics

In the DQFM treatment we assume the existence of a fracture quantum and correspondingly the energy balance has to be satisfied during a time quantum (a time to failure) connected to the time to produce a fracture quantum, which is finite as a consequence of the finite crack speed. Thus, the quantization (one might also call it "discretization") is assumed in both space and time. The energy balance in the continuum space-time is "virtual" and becomes real only for the real formation of a fracture quantum. The classical energy balance is thus rewritten as a quantum action balance, i.e., as: $\int_{t-\Delta t}^{t} \Delta(W + T + \Omega) dt = 0$, where the finite difference is related to the quantized crack advancement; thus, it is equivalent to $\frac{1}{\Delta t} \int_{t-\Delta t}^{t} \frac{\Delta}{\Delta A}(W + T + \Omega) dt = 0$, where $\Delta A$ and $\Delta t$ are the time and fracture quanta (the finite variations in the integral are with respect to the crack surface area) and $\frac{1}{\Delta t} \int_{t-\Delta t}^{t} \frac{\Delta}{\Delta A} \Omega dt \equiv G_{dC}$. Defining the dynamic quantized energy release rate as:

$$G_d^* \equiv \left\langle \left\langle G_d \right\rangle_A^{A+\Delta A} \right\rangle_{t-\Delta t}^t = -\left\langle \Delta(W+T)/\Delta A \right\rangle_{t-\Delta t}^t, \tag{7}$$

the criterion $G_d^* = G_{dC}$ describes the quantized crack propagation under time-dependent loading conditions (here $\langle f \rangle_{x1}^{x2}$ represents the mean value of $f$ in the interval $(x1, x2)$). The quasi-static condition corresponds to QFM and becomes $G^* = G_C$ [6], where:

$$G^* \equiv \langle G \rangle_A^{A+\Delta A} = -\Delta W/\Delta A, \tag{8}$$

is the (static) quantized energy release rate (we note that if $G_d(l,t,v) = g(v)G_d(l,t,0)$ is valid, then $G_d^*(l,t,v) = g(v)G_d^*(l,t,0)$). Correspondingly the dynamic quantized Irwin's correlation is:

$$G_d^* = \frac{A_I(v)K_{dI}^{*2}}{E'} + \frac{A_{II}(v)K_{dII}^{*2}}{E'} + \frac{1+\nu}{E} A_{III}(v)K_{dIII}^{*2}, \tag{9}$$

where $K_{dI,II,III}^*$ are the dynamic quantized stress-intensity factors ($K_{dI,II,III} > 0$) defined by:

$$K_{dI,II,III}^* \equiv \sqrt{\left\langle \left\langle K_{dI,II,III}^2 \right\rangle_A^{A+\Delta A} \right\rangle_{t-\Delta t}^t}. \tag{10}$$

Thus, the incipient crack propagation in the quasi-static quantized based treatment [6] is $K_{I,II,III}^* = K_{I,II,IIIC}$, where:

$$K^*_{I,II,III} \equiv \sqrt{\left\langle K^2_{I,II,III} \right\rangle_A^{A+\Delta A}}, \tag{11}$$

whereas in the general dynamic treatment of DQFM it is:

$$G^*_d = G_{dC} \quad \text{or} \quad K^*_{dI,II,III} = K_{dI,II,IIIC} \quad \text{(DQFM)}. \tag{12}$$

The criterion $G^*_d = G_{dC}$ can be used also for mixed mode crack propagation (the crack will propagate in the direction of the maximum energy release rate, see [6]), whereas the criterion $K^*_{dI,II,III} = K_{dI,II,IIIC}$ is valid only for pure crack propagation modes.

In contrast to DFM, to apply DQFM for predicting the strength (or time to failure) of solids $G_{dC} \equiv G_C$ and $K_{dI,II,IIIC} \equiv K_{I,II,IIIC}$ (for $v$=0 $g_C = k_{I,II,IIIC} = 1$) and thus, an *ad hoc* dynamic fracture initiation toughness does not have to be postulated. If $K_{dI,II,II} = k_{I,II,III}(v) K_{I,II,III}$, then the expressions for $K_{I,II,III}$ are reported for hundreds of cases in the stress-intensity factors handbooks [14,15]. Note that, as the well-known Neuber-Novozhilov [16,17] approach, our theory is still based on continuum linear elasticity. There is, in fact, a perfect parallelism between them. Thus, the physical meaning of the stress-intensity factors is obvious. However, we note that the assumption of a discrete crack advancement –intrinsically introducing and quantifying some classical "nonlinear" effects such as the *R*-curve behaviour– seems to be a powerful tool for treating fracture in discrete lattices.

Eqs. (1-12) define DQFM completely, predicting the failure strength $\sigma_f$, the time to failure $t_f$ and the dynamic crack tip evolution $v(t)$, for general time-dependent loading conditions $\sigma = \sigma(t)$, assuming the energy release rate to be quantized in both space and time. DQFM treats any defect size and shape (as QFM, see [6]) and loading rate. It is evident that interesting limit conditions for DQFM are (we now omit the symbols *I,II,III*):

$$\text{DQFM}: \begin{cases} \sigma = \sigma(t): G_d^* = G_{dC} \equiv g_C G_C, K_d^* = K_{dC} \equiv k_C K_C \to v(t) \\ \sigma = \sigma(t), v = 0: G_d^* = G_C, K_d^* = K_C \to \sigma_f, t_f \\ \to \text{QFM}: \sigma \neq \sigma(t) \to \Delta t = 0; v = 0: G^* = G_C, K^* = K_C \to \sigma_f \\ \to \text{DFM}: \Delta t = 0; \Delta A = 0: G_d = G_{dC}, K_d = K_{dC} \\ \to \text{LEFM}: \Delta t = 0; \Delta A = 0; v = 0: G = G_C, K = K_C \end{cases}$$

## 4. The tensional analog of the action-based DQFM

Let us assume a fracture quantum of length *a* (e.g., $\Delta A \equiv ah$ in a plate having height *h*). The time quantum is expected to be of the order of $\Delta t \approx a/v$. Indicating with $\sigma_y$ the stress acting at the tip (placed at *x*=0) of a defect, the stress analog of DQFM for the strength prediction must be written as:

$$\sigma_d^* \equiv \frac{1}{a\Delta t} \int_{t-\Delta t}^{t} \int_0^a \sigma_y(x,t) \mathrm{d}x \mathrm{d}t \geq \sigma_C. \tag{13}$$

This crack propagation criterion has been formulated as the dynamic extension of the Neuber-Novozhilov criterion [16,17] and successfully applied in the study of dynamic crack propagation under high loading rate conditions by Morozov, Petrov and Utkin [18] (see also [19]) that consider $\Delta t$ as an incubation time to failure, a characteristic relaxation time upon micro-fracture of a material. The analogy with DQFM for predicting the structural strength and time to failure for pure crack modes is evident by rewriting the DQFM criterion for crack propagation as:

$$K_d^* = \sqrt{\frac{1}{a\Delta t} \int_{t-\Delta t}^{t} \int_0^a K_d^2(x,t) \mathrm{d}x \mathrm{d}t} \geq K_C. \qquad (14)$$

**5. The equation of the dynamic *R*-curve and of the dynamic fracture resistance**

For the continuum approach, the measured (superscript (*m*)) dynamic fracture energy (which is undefined in the classical treatment) $G_C^{(m)}$ is a function (the so-called *R-curve*) of geometry, length and crack/loading speed (and it is thus not a material property, see [20]). To obtain the same predictions of DQFM by applying the classical DFM, one is forced to assume an unrealistic dynamic resistance curve (thus not a material property, e.g., a function of the crack length, structural size and shape, time to failure, and so on…) $G_{dC} \to G_{dC}^{(m)} \equiv R$. By comparing the DQFM and DFM treatments, we find:

$$R = g_C G_C + G_d - G_d^*; \qquad (15)$$

Accordingly, if the continuum approach is used in the stress-intensity factor treatment, one would measure (subscript (*m*)) a dynamic fracture toughness:

$$K_{dC}^{(m)} = k_C K_C + K_d - K_d^*, \qquad (16)$$

observed to be different from $K_C$ (or equivalently $R$ from $G_C$) also at the incipient propagation (where $v=0$ and $g_C = k_C = 1$) (see [20]). In contrast to DQFM, continuum approaches are unable to explain why at the incipient crack propagation $R$ is different from $G_C$, or $K_{dC}^{(m)}$ from $K_C$. As we are going to show, DQFM is able to quantitatively predict such a fictitious discrepancy.

**6. Simple examples: strength, time to failure, and crack tip equation**

We consider the Griffith's case (a) of a linear elastic infinite plate in tension, of uniform thickness $h$, with a crack initial length $2l_0$ orthogonal to the applied far field (crack opening Mode I). The material is described by the fracture toughness $K_{IC}$ and the fracture quantum at the considered size-scale $\Delta A \equiv ah$. For this case, as it is well known, $K_I(l) = \sigma\sqrt{\pi l}$, where $\sigma$ is the applied time-independent far field stress. In this first simple case we consider a time-independent stress-intensity factor. According to DQFM (or QFM) the failure strength is:

$$\sigma_f = \frac{K_{IC}}{\sqrt{\pi(l_0 + a/2)}}. \qquad (17)$$

Note that in this case the tensional analog (the static case of eq. (13) and developed by Neuber-Novozhilov [16,17]), considering the complete stress field at the tip of a crack, gives the

*identical* result but in a less simple way, as demonstrated in [21] in which basically an extensive data fitting of QFM to larger size scales experiments is successfully presented. Inverting eq. (17), the fracture quantum can be estimated from the mechanical properties at a given size-scale, $\sigma_C = \sigma_f(l_0/a \to 0)$ and $K_{IC}$, as $a = 2K_{IC}^2/(\pi\sigma_C^2)$.

Let us assume in this example for the sake of simplicity $g_C \approx 1$ and $G_d \approx g(v)G$. The dynamic evolution under the constant applied stress $\sigma_f$ causing the initiation of the crack propagation, is predicted by DQFM as:

$$\frac{v}{c_R} = 1 - \frac{l_0 + a/2}{l + a/2}. \qquad (18)$$

According to eq. (18) the Griffith's crack is predicted to be unstable. The time evolution of the crack tip could be obtained by solving the differential equation (18), where $v = dl/dt$. For LEFM and DFM, the predictions of eqs. (17) and (18) would be the same if the fracture quantum is assumed to be negligible. As expected, the results of the quantized approach tend to the classical values if the continuum hypothesis $a/l, a/l_0 \to 0$ is made. The corresponding result for $\sigma_f$, in contrast to eq. (17), would be without meaning for $l_0 \to 0$, predicting an infinite ideal strength. In contrast, if the fracture quantum corresponds to the atomic size, the ideal strength $\sigma_C = \sigma_f(l_0/a \to 0)$ predicted by eq. (17) is identical to Orowan's prediction [22] if multiplied by a factor of $\sqrt{\pi/4} \approx 1$, as discussed also in reference [6]. Note that the experimentally observed asymptote of $v/c_R < 1$ for $l/l_0 \to \infty$ can be explained by generation of secondary cracks from the tip of the predominant one [23] and thus can not be deduced from the pure Griffith's case (i.e., eq. (18)), in which no interacting cracks are considered.

By applying eq. (15), and assuming $v=0$, we find the expression of the (static) *R-curve* as:

$$R = \frac{G_C}{1 + a/(2l_0)}, \qquad (19)$$

thus, as expected (see [20]), $R$ increases and tends to $G_C$ for crack length tending to infinite size scales.

Taking into account the blunting of the crack tip (for example, due to the opening of two dislocations at the tip, see [23]), we have to make the substitution $G_C \to G_C(1 + \rho_0/2a)$ in the previous equations [6]. Eqs. (17) and (18) would become:

$$\sigma_f = K_{IC}\sqrt{\frac{1 + \rho_0/2a}{\pi(l_0 + a/2)}} = \sigma_C\sqrt{\frac{1 + \rho_0/2a}{1 + 2l_0/a}}, \qquad (20)$$

$$\frac{v}{c_R} = 1 - \frac{l_0 + a/2}{l + a/2}\frac{1 + \rho/2a}{1 + \rho_0/2a}, \qquad (21)$$

where $\rho$ and $\rho_0$ are the tip radii of the cracks of length $l$ and $l_0$ respectively, and $\sigma_C = \sigma_f(l_0 = 0, \rho_0 = 0)$. Note that, if the continuum hypothesis is made ($a/l_0, a/\rho_0 \to 0$), eq. (20) yields practically the same result as the classical tensional approach (maximum stress equal to material strength), for which the stress concentration $\sigma_C/\sigma_f$ is $1 + 2\sqrt{l_0/\rho_0} \approx 2\sqrt{l_0/\rho_0}$ (small radii) as given by the Theory of Elasticity. Thus, eq. (20) represents the link between concentration and intensification factors. It predicts a finite strength that is size-dependent (in contrast with the continuum tensional approach coupled with the Theory of Elasticity) for geometrical self-similar defects in an infinite plate.

We consider a complementary case (b), a stationary crack, for which the stress-intensity factor is independent from the crack length. A semi-infinite crack in an otherwise unbounded body is considered. The body is initially stress free and at rest. At time $t=0$ a self-balanced antiplane shear $\tau$ begins to act on the crack faces. In this case, as it is well known, $K_{III}(t) = 2\tau\sqrt{\dfrac{2c_s t}{\pi}}$ (see [5]). The time to failure $t_f$ is predicted by DQFM to satisfy the following relationship ($t_f > \Delta t$):

$$2\tau\sqrt{\frac{2c_S}{\pi}} = \frac{K_{IIIC}}{\sqrt{t_f - \Delta t/2}}. \qquad (22)$$

Note that, according to our time quantization, a minimum time to failure exists and it must be of the order of $t_{f\min} \approx \Delta t$. This could represent an additional physical meaning of the time quantum. On the other hand, by applying DFM, we obtain the same result of eq. (22) if the time quantum is neglected. The equation of the $R$-curve, according to eq. (15), for $v=0$ is:

$$R = \frac{G_C}{1 - \Delta t/(2t_f)}, \qquad (23)$$

and it decreases, tending to $G_C$ when time to failure tends to infinity. Since $t_{f\min} \approx \Delta t$, for this case, the measured dynamic fracture initiation energy is predicted approximately to be twice its static value by varying the time to failure within several orders of magnitude. If one applies the classical DFM, then according to DQFM an "apparent" dynamic resistance doubled with respect to the static value is obtained. This behaviour is observed experimentally, as we will discuss in the following.

**7. Strength and crack speed forbidden bands**

Let us reconsider the Griffith's case. As stated for DQFM the crack length is quantized, and so $2l_0 = n_0 a$ and $2l = na$, from which the quantization of the strength and crack speed can be deduced. For the Griffith's case, from eqs. (20) and (21) and assuming a blunt crack due to adjacent vacancies (i.e., $2\rho_0 \approx a$), and time-independent blunt tips ($\rho_0 \approx \rho$) we have:

$$\sigma_f = \sigma_C \sqrt{\frac{5/4}{1+n_0}}, \quad n_0 > 0, \qquad (24)$$

$$\frac{v}{c_R} = 1 - \frac{1+n_0}{1+n}, \quad n \geq n_0 \geq 0. \qquad (25)$$

The first (and largest) forbidden strength band (between $\sigma_C$ and the prediction for $n_0 = 1$) is in the range $(1-0.8)\sigma_C$; atomistic simulations of two-dimensional lattices with adjacent vacancies quantitatively agree with such forbidden bands (see [6]); for example just one vacancy is expected to reduce the strength by a factor of ~20%. Furthermore, DQFM derives forbidden crack speed bands. Starting from our simple assumption of $g_C = 1$, $G_d = g(v)G$, the first band (between the predictions for $n = n_0, n_0 +1$) is in the range $(0 - (1+n_0)/(2+n_0))v/c_R$; the smallest ($n_0 = 0$), corresponding to the crack initiation from a plain specimen, is $(0 - 0.5)v/c_R$. Atomistic simulations of crack evolution in two-dimensional lattices qualitatively show such forbidden bands (see [23]). They imply hysteretic crack motion (hysteretic cycles in the crack speed versus applied load curves), known as "lattice trapping". Postulated since the early 1970s, it has been observed numerically but never experimentally (see [23]). Due to the large size scale of the experiments (implying large pre-existing cracks, i.e., large values of $n_0$) such strength and crack speed forbidden bands are difficult to observe; on the other hand, in contrast to continuum based theories, DQFM implies such strength and crack speed "quantizations", which may be detectable in nanoscale experiments.

## 8. Quantum Mechanics analogy

DQFM presents an analogy with quantum mechanics as a consequence of the *action quantum* in each: $G_C \Delta A \Delta t$, and $\hbar$, respectively. For example, in reducing the absolute temperature of a solid, classical physics predicts its internal energy approaches zero (unstressed specimen). In the quantum mechanical treatment it approaches a finite value (the zero point energy). For elevated temperatures the two treatments yield essentially identical energies (Correspondence Principle). The analogy is thus between the curve *internal energy vs. temperature* and the *inverse of the strength vs. crack length/life-time* as given by DQFM (and the Correspondence Principle for

DQFM is with respect to LEFM and DFM respectively). Other analogies can be found in the discussion by Petrov [19].

## 9. Static resistance: an application for predicting the fracture strength of defective nanotubes

The strength and fracture of the outer shell of multi-walled carbon nanotubes (MWCNTs) is reported in [24]. The tensile strengths of this outer shell for 19 individual MWCNTs were measured with a *nanostressing stage* having two opposing atomic force microscope (AFM) tips, and operated in a scanning electron microscope (SEM). This tensile strength ranged from 11 to 63 GPa for the set of 19 MWCNTs that were loaded (in particular, values of 63, 43, 39, 37, 37, 35, 34, 28, 26, 24, 24, 21, 20, 20, 19, 18, 18, 12, 11 GPa were measured).

From such experimental results, distinct clusters about a series of decreasing values of strength, with the maximum 63 GPa, and other values at 43, and in the ranges 36-37, 25-26, 19-20 and 11-12 GPa, were observed. The highest measured value of 63 GPa is lower than the ideal tensile strength of small diameter carbon nanotubes (CNTs), recently obtained with density functional theory (DFT, [25]). If the fracture quantum is assumed to be the distance between two adjacent broken chemical bonds, i.e., $a \approx \sqrt{3}r_0$, with $r_0 \approx 1.42\,\text{Å}$ and adjacent vacancies are considered, i.e., $2l_0 = n_0 a$ in eq. (19), the predicted strength quantizations for $n_0 = 2,4,6,8$ (with $\rho_0 \approx 0.8a \approx 2.0\,\text{Å}$, shown in Figure 1a) are in close agreement with Molecular Mechanics (MM) calculations [26], see [6]. The result is that the strength is strongly reduced by the presence of the nanoflaws. This contradicts the statement that *materials become insensitive to flaws at nanoscale*, as claimed in the title of ref. [27].

The initial crack speed, for the different cases of $n_0 = 2,4,6,8$, would be estimated to be respectively of $v/c_R = 3/4, 5/6, 7/8, 9/10$ (but we note that such estimations refer to the overly simplified assumption of $g_C = 1$, $G_d = g(v)G$).

Different kinds of defects, such as holes, might be more stable than crack-like defects at the nanoscale [28, 29]. Nanotubes with "pinhole" defects have been recently investigated -by MD simulation [28]. In this context, quantum mechanical calculations using DFT theory, semiempirical methods and MM simulations have been recently performed [29]. The results of the atomistic simulations [29] were compared with QFM in [6], with close agreement (nano-holes are as shown in Figure 1b).

We assumed (see ref. [6] for details) for such (large) nanotubes as the outer shell in the 19 MWCNTs (the diameter varied from 20 to 40 nm), that the cross section reduction due to the presence of defects was negligible. It is interesting that, enforcing this constraint means that the strength tends asymptotically to a finite value (1/3.36 the strength of the structure without the hole according to QFM, in agreement with MM simulations [6]); this is however still larger than the smallest values experimentally measured. Perhaps (i) sharper defects as discussed above, or (ii) larger holes (breakdown of the assumption of no reduction in cross section), or (iii) "small" holes satisfying the cross-section constraint, but that are close and thus causing a greater stress concentration between them then would be the case if they were isolated, are all possible reasons for strength values as low as 11 GPa.

An additional intermediate type of defect was numerically treated in [30]; it corresponds to an elliptical hole with size that we define by an index *i*. Such a defect was not treated in [6].

Starting from a hole obtained removing 6 atoms at the vertexes of an hexagon ($i$=1) the other defects corresponding to larger sizes and indexes $i$ are obtained removing the lateral four carbon atoms at each blunt tip (see Table 1). The comparison between QFM (the case of pinhole defect $i$=1 is treated in [6], whereas here we simply consider $\rho_0 \approx 0.9a$ and $l \approx (2i-1)r_0$, $i$=2,3,4,5, in eq (20)) and atomistic simulations [30] is reported in Table 1. However, we note that the numerically observed strength asymptote for increasing crack length is "unexpected" and thus unclear, since for such a case the crack becomes macroscopic and classical fracture mechanics would suggest the strength decreasing to zero.

For an ideal strength for the experimentally investigated MWCNTs [25] assumed to be 93.5 GPa (as computed in [26]) the corresponding strength for an $i$=1 defect is 64 GPa (compared to the measured value of 63 GPa), for an $i$=3 defect it is 43 GPa (in agreement with the measured value), for an $i$=4 defect it is 37 GPa (against the measured value of 39 GPa), for $i$=5 defect is 34 GPa (against the measured values of 35 and 34 GPa), for $i$=6 defect is 30 GPa (against the measured values of 28 GPa), and so on. This could represent a more plausible scenario (since elliptical holes are chemically more stable than crack-like defects) compared to the assumed linear defects (and circular holes) that were discussed in [6]. LEFM cannot treat blunt, or short short cracks, or holes; DQFM/QFM can.

# 10. Dynamic resistance: an application for predicting the time to failure of 2024-T3 aircraft aluminium alloy

In this section we refer to the experimental work discussed in [31]. In [32] eq. (13) is applied with respect to time considering only the asymptotic part of the stress field, thus $\sigma_y(t) \propto K_I(t)$ (and $\sigma_C \propto K_{IC}$) to rationalize some of the experimental results [31]. The expression of the dynamic stress-intensity factor as applied to the experiments is obtained by considering an infinite elastic plane containing a semi-infinite crack subjected at $t=0$ to a linearly increasing impact load [32]. The measured dynamic fracture initiation toughness can be obtained according to eqs. (14) and (16), where here $v=0$, so that $k_C = 1$. According to DQFM, we find the result as:

$$\frac{K_{dIC}^{(m)}}{K_{IC}} = \frac{2 t_f^{3/2} \sqrt{\Delta t}}{\sqrt{t_f^4 - (t_f - \Delta t)^4}}$$

, thus as a function of the time to failure. This function is reported as the solid line in Fig. 2 assuming $\Delta t = 50 \mu s$, whereas the dots refer to the DFQM tensional analog of eq. (13) fitted with good agreement with the experimental results [31] assuming $\Delta t = 40 \mu s$: see [32] for details. Since the two criteria are different, different values for $\Delta t$ were expected: however we note that the two values are close. In addition, we note that from the DQFM prediction, a minimum time to failure $t_f \approx \Delta t$ is expected, corresponding to a dynamic fracture initiation toughness of $K_{dIC}^{(m)} \approx 2 K_{IC}$. Note that in [31] the authors report the observation of "a minimum time necessary to initiate crack growth", of the order of $t_f \approx 75 \mu s$, and a dynamic fracture initiation toughness that "reveals an increase of a factor of ~2, as the loading rate increases by seven orders of magnitude" (or as the time to failure decreases). DQFM is in good agreement with the experimental data and with the criterion of eq. (13).

DFM is unable to explain such observations (e.g., the apparent variation of the dynamic fracture initiation toughness). DQFM offers explanations and it is of interest to see further experimental data with which it could be assessed.

## 11. Concluding remarks

DQFM has been presented and used to describe the study of the strength and time to failure of solids, as well as for the study of the time evolution of the crack tip, also at nanoscale.

For example DQFM can be used as a tool in the design of the *Space Elevator* cable based on carbon nanotubes (since no experiments or numerical atomistic simulations could be used for such a large scale). For example, assuming in such a cable large holes (very likely as a consequence of its large size), DQFM predicts an asymptotic limit value of $\sigma_C/\sigma = 3.36$; thus, assuming the ideal nanotube strength of $\sigma_C = 93.5\text{GPa}$ [26], we obtain a value of $28\text{GPa}$. Considering in addition the actual nanotube cross-section area, we estimate the ratio between the resistant and apparent cross-section area as $\eta \approx \dfrac{\pi(R_e^2 - R_i^2)}{2\sqrt{3}(R_e + d/2)^2}$, with $R_i, R_e$ the inner and outer nanotube radii and $d \approx 0.334\text{nm}$; for $R_i \approx 0$ and $R_e >> d$, $\eta \approx \dfrac{\pi}{2\sqrt{3}} \approx 0.9$ and the prediction for the cable strength is $\eta \cdot 28\text{GPa} \approx 25\text{GPa}$. Since more critical defects could also be present in such a cable, this can be considered a statistically plausible upper bound for its strength, thus much lower than the ideal strength of nanotubes.

**Acknowledgements**

The authors would like to thank J. Weertmann, T. Belytschko and Y. Petrov for commenting on the manuscript. RSR appreciates support from the NSF grant no. 0200797 "Mechanics of Nanoropes" (Ken Chong and Oscar Dillon, program managers), from the ONR grant no. N00014-02-1-0870 "Mechanics of Nanostructures" and from the NASA University Research, Engineering and Technology Institute on Bio Inspired Materials (BIMat) under award No. NCC-1-02037.

# FIGURE CAPTIONS

Figure 1:

(a) Atomic *n*-vacancy defects and short blunt cracks used for predicting the strength by QFM in [6]: the crack length was imposed as *na*, whereas the blunt tip radius was chosen to fit the ideal nanotube strength; as shown, the blunt tip radius appears to be reasonable. The fracture quantum length *a* is the distance between two adjacent parallel C-C bonds.

(b) Hole used for predicting strength by QFM in [6]. The fracture quantum is again fixed as identical to the length between two adjacent parallel C-C bonds. (Note that "opened bonds" are not shown in the figure, and for this reason it appears as if the smallest circles seem to "underestimate" the defect size (in reality, they do not)).

Figure 2:

Dynamic fracture initiation toughness over (static) fracture toughness, as a function of time to failure for 2024-T3 aircraft aluminum alloy. Solid-line obtained by DQFM ($\Delta t = 50 \mu s$); the dots are from the tensional analog of DQFM shown to be in good agreement with the experiments when fitted using $\Delta t = 40 \mu s$ (see [32]).

# FIGURES

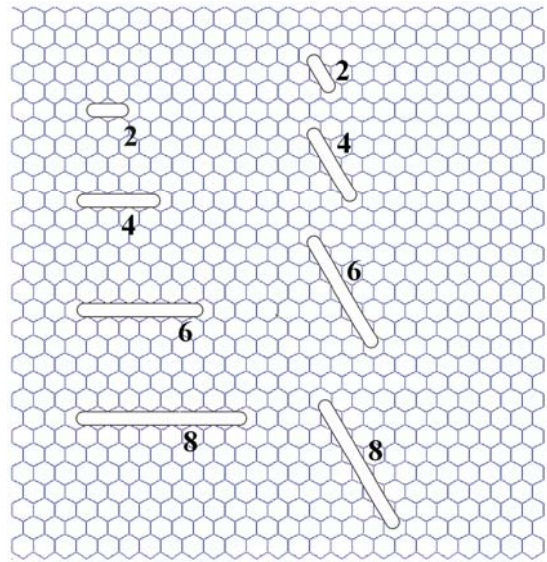

(a)

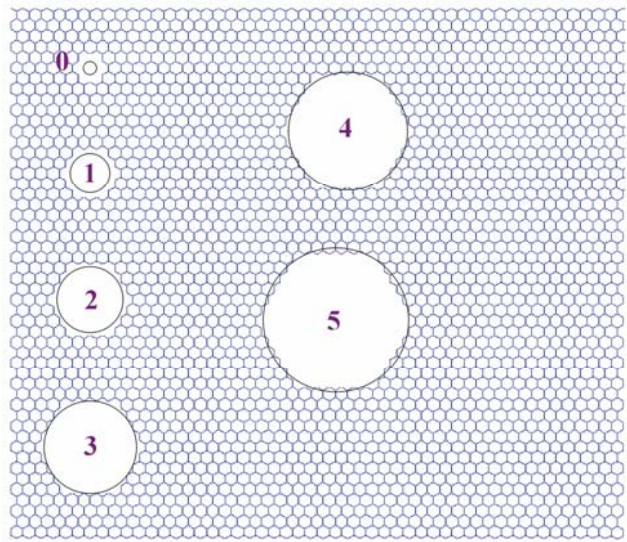

(b)

*Figure 1*

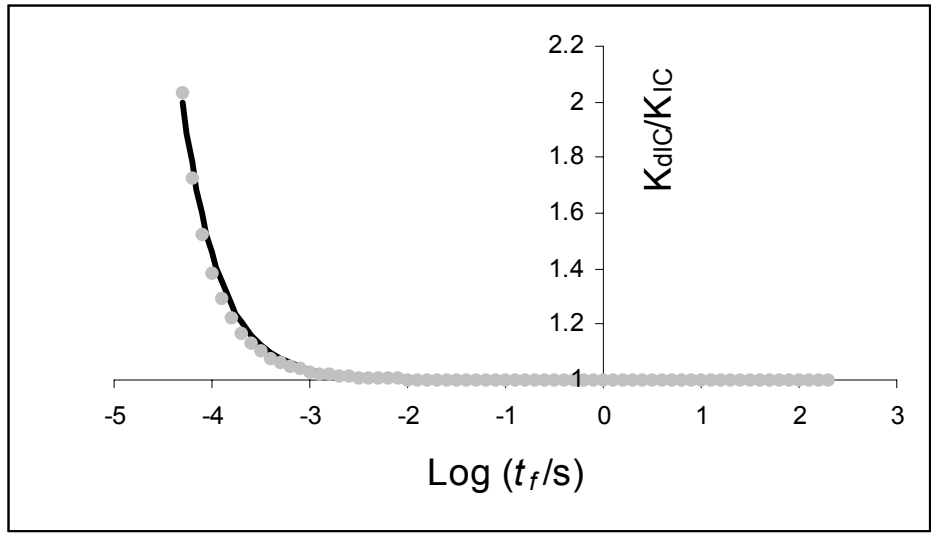

*Figure 2*

# TABLE CAPTIONS

Table 1: Comparison between fracture strengths of a (50,0) nanotube, obtained by MM and by QFM, with elliptical holes of size *i* (the graph shows the example of *i*=1 and the atoms (in black) that would be removed to generate the *i*=2 defect).

# TABLES

| $\sigma/\sigma_C$ | *i*=1 | *i*=2 | *i*=3 | *i*=4 | *i*=5 | *i*=6 | 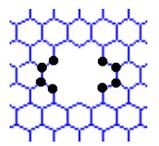 |
|---|---|---|---|---|---|---|---|
| Theo. | 0.68 | 0.57 | 0.46 | 0.40 | 0.36 | 0.32 | |
| Num. (50,0) | 0.64 | 0.51 | 0.44 | 0.40 | 0.37 | 0.34 | |

*Table 1*